\documentclass[
aps,
twocolumn,
prd,
superscriptaddress,
letterpaper,
nofootinbib,
longbibliography]{revtex4-2}
\usepackage{amssymb}
\usepackage{physics}
\usepackage{mathtools}
\usepackage{upgreek}
\usepackage{placeins}
\usepackage{booktabs}
\usepackage{subcaption}
\usepackage{lipsum}
\AtBeginDocument{
	\heavyrulewidth=.08em
	\lightrulewidth=.05em
	\cmidrulewidth=.03em
	\belowrulesep=.65ex
	\belowbottomsep=0pt 
	\aboverulesep=.4ex
	\abovetopsep=0pt
	\cmidrulesep=\doublerulesep
	\cmidrulekern=.5em
	\defaultaddspace=.5em
}
\usepackage{lineno}  
\usepackage[justification=Justified,format=plain]{caption}

\usepackage{txfonts}
\usepackage{bm}
\usepackage{placeins}
\usepackage{setspace}

\usepackage{tikz}
\usetikzlibrary{fit}
\usetikzlibrary{calc,positioning,shapes.misc}
\usetikzlibrary{arrows.meta,trees,shadings}
\usetikzlibrary{shapes.geometric}
\usetikzlibrary{intersections}

\newcommand{\laserOneTwo}{LA 12}
\newcommand{\laserOneThree}{LA 13}
\newcommand{\laserTwoOne}{LA 21}
\newcommand{\laserTwoThree}{LA 23}
\newcommand{\laserThreeOne}{LA 31}
\newcommand{\laserThreeTwo}{LA 32}

\newcommand{\spacecraftSeparation}{50mm}
\newcommand{\spacecraftRadius}{11mm}

\usepackage[colorlinks=true]{hyperref}

\begin{document}

\title{Complete Catalog of Laser Locking Configurations for LISA}

\author{Gerhard Heinzel}
\email{gerhard.heinzel@aei.mpg.de}
\affiliation{Max Planck Institute for Gravitational Physics (Albert Einstein Institute), D-30167 Hannover, Germany\\
Leibniz Universit\"at Hannover, D-30167 Hannover, Germany}

\author{Javier \'Alvarez-Vizoso}
\affiliation{Max Planck Institute for Gravitational Physics (Albert Einstein Institute), D-30167 Hannover, Germany\\
Leibniz Universit\"at Hannover, D-30167 Hannover, Germany}

\author{Miguel Dovale-\'Alvarez}
\affiliation{Max Planck Institute for Gravitational Physics (Albert Einstein Institute), D-30167 Hannover, Germany\\
Leibniz Universit\"at Hannover, D-30167 Hannover, Germany}


\begin{abstract}
The Laser Interferometer Space Antenna (LISA) will enable direct observations of low-frequency gravitational waves, offering unprecedented insight into astrophysical and cosmological phenomena. LISA's heterodyne interferometric measurement system requires phase-locking five of its six onboard lasers with tunable frequency offsets to ensure that all beatnotes remain within the metrology system's operational range, despite Doppler-induced frequency shifts. The selection of these offset frequencies---collectively forming a frequency plan---is a complex optimization problem constrained by the spacecraft's orbital dynamics and instrument limitations. While previous work established an algorithmic solution for deriving time-dependent frequency plans, this study takes a complementary approach by systematically analyzing and cataloging all possible laser locking configurations. We present an automated method to explore, validate, and classify viable locking schemes, identifying 36 unique non-frequency-swapping configurations and 72 additional frequency-swapping configurations for an arbitrary choice of primary laser. This exhaustive classification provides a foundation for frequency planning across the full range of operational scenarios.
\end{abstract}

\maketitle

\section{Introduction} \label{section:introduction}

The advent of gravitational-wave astronomy has unveiled a previously inaccessible spectrum of astrophysical phenomena, enabling transformative discoveries through ground-based detectors such as LIGO and Virgo~\cite{Miller2019, Collaboration2015, AdvancedVirgo15}. These facilities have detected hundreds of gravitational-wave events, including from merging binary black holes~\cite{Abbott2016}, binary neutron stars~\cite{Abbott2017}, and neutron star-black hole mergers~\cite{Abbott2021}. However, their sensitivity is fundamentally limited at low frequencies by seismic and control noise~\cite{Cahillane2022}. To overcome this limitation, the Laser Interferometer Space Antenna (LISA) will operate in space, allowing the detection of gravitational waves in the millihertz regime, where signals from supermassive black hole mergers and early universe relics are expected to be most prominent~\cite{Amaro2023}.

LISA consists of three spacecraft (SC) forming an approximately equilateral triangle with 2.5-million-kilometer arm lengths. Each spacecraft hosts two lasers, establishing six inter-spacecraft laser links. Unlike terrestrial interferometers, where reflections from mirrors form standing wave cavities, LISA relies on a heterodyne detection scheme where phase measurements are derived from beatnotes between weak incoming and strong local laser fields. Since the inter-spacecraft distances vary due to orbital dynamics, the laser beams experience Doppler shifts of their frequency up to $\pm 8$\,MHz, necessitating an adaptive laser frequency locking scheme to keep the beatnotes within the phasemeter's operational range (e.g., 5–25\,MHz).

In this locking scheme, one laser is designated as the primary and stabilized to an ultra-stable optical cavity. The remaining five lasers are phase-locked with programmable frequency offsets to ensure proper heterodyne detection. The selection of these offsets forms the frequency plan, which must be dynamically adjusted throughout the mission to prevent beatnotes from drifting outside the allowed range. In previous work~\cite{Heinzel2024}, an optimization framework was developed to generate time-dependent frequency plans satisfying these constraints, ensuring uninterrupted scientific operation.

Here, we extend that work by systematically analyzing the full space of laser locking configurations. Specifically, we develop an automated method to exhaustively explore and validate phase-locking schemes by symbolically manipulating the governing frequency equations. We distinguish between frequency swapping and non-swapping schemes, depending on the design topology of the long arm interferometer. This analysis reveals that for any given choice of primary laser, there exist 6 distinct non-frequency-swapping configurations and 12 frequency-swapping configurations. In total, across all possible primary laser assignments, there are 36 and 72 configurations, respectively. The resulting catalog provides a comprehensive classification of feasible locking setups, offering a crucial tool for frequency planning across any possible scenario throughout the mission lifetime.

The paper is structured as follows. Section~\ref{section:method} describes the methodology used to derive and validate locking configurations, including a step-by-step example of the N3-L32 configuration. Section~\ref{section:results} presents a catalog of the six fundamental locking configurations for the case without frequency swaps. It also details the cataloging method we developed, where different laser locking schemes are sorted in ascending order of interferometric complexity. The complete catalog, including frequency-swap configurations, is provided as supplementary material. Finally, Section~\ref{section:conclusion} summarizes our findings and their implications for LISA's frequency planning and interferometric operation.

\section{Method} \label{section:method}

\begin{figure*}
\centering
\includegraphics{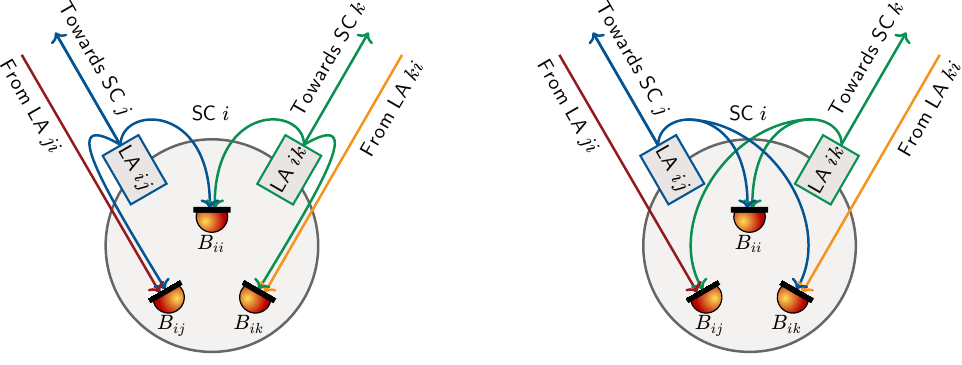}
\caption{Illustration of frequency swapping. In the non-swap configuration (left), each laser serves both as the transmit beam along an arm and as the local oscillator for the weak incoming beam in the same direction. In the frequency-swap configuration (right), the local oscillator for each received beam is swapped to the laser pointing along the opposite arm, thereby shifting the frequency of any spurious back-scatter away from the science beatnote. This mitigates susceptibility to small vector noise but introduces increased complexity. The arrows highlight the routing of the optical beams and the reassignment of local oscillator roles across the two architectures.}
\label{figure:swapping}
\end{figure*}

\subsection{Frequency swapping}

There are two lasers on each spacecraft, one transmitted along each long arm, and three different beatnote frequencies. The first beatnote frequency is the one formed between the two local lasers, and is needed as a stable phase reference for time delay interferometry. Moreover, the beatnote of the two local lasers is also formed with light that has reflected off the local test masses, and is used for test mass displacement readout. Therefore, the so called reference and test mass interferometers (and possibly other auxiliary interferometers) both use beatnotes of the same frequency.

The other two beatnote frequencies in each spacecraft involve the weak light received from the remote spacecraft along each arm. This light is interfered with one of the two local lasers (called ``local oscillator'') to produce detectable beatnotes in what is referred to as the long arm or science interferometers.

Frequency swapping is a design topology that was considered to mitigate the effect of scattered light returning from the transmit telescope in the same optical axis as the weak received beam. The two design topologies are illustrated in Figure~\ref{figure:swapping}. In the conventional, ``non-swap'' configuration, the laser transmitted along one arm is also used as local oscillator for the received light in that same arm. Conversely, in the frequency-swap configuration, the laser used as local oscillator for each weak received beam is the laser transmitted along the opposite arm. Without frequency swap, that scattered light is at the same frequency as the local oscillator used to detect the incoming weak light. If its phase changes, it can produce what is known as ``small vector noise''. With frequency-swap, the science interferometers are insensitive to direct back-scatter of the transmitted light from the telescope, as that local laser does not participate in the measurement.

The impact of back-scattering in the science interferometers is difficult to predict, as it depends on the amount of scattered light, its ability to interfere with the weak received beam, and its phase stability. In the frequency-swap scheme, the offending beatnote is shifted to the frequency of the local reference interferometer. The amplitude of this spurious beatnote is expected to have significantly higher amplitude than the desired signal. Early studies rejected frequency swapping due to concerns that this would reduce the dynamic range of the phase measurement chain, particularly if low-bit A/D converters were used. However, with the availability of 14-bit converters, this limitation is now considered less critical. Nevertheless, the frequency-swap scheme introduces additional complexity in the frequency distribution system, and other factors, including redundancy, must be considered. The present baseline for LISA is to use the non-swap scheme.

\begin{figure*}
\centering  
\begin{minipage}[t]{0.48\textwidth}  
\centering  
\begin{tikzpicture}
\definecolor{dkblue}{rgb} {0.00,0.33,0.68}
\definecolor{dkred}{rgb} {0.80,0.20,0.00}
\definecolor{lblue}{rgb} {0.00,0.60,0.95}
\definecolor{lred}{rgb} {0.95,0.30,0.00}
\tikzset{spacecraft/.style={circle, minimum width = 24mm, draw=black!60, fill=gray!10, very thick}}
\tikzset{laser/.style={rectangle, minimum width=10mm, minimum height=4mm, fill=red!20, draw=red!80, thick}}
\tikzset{masterlaser/.style={rectangle, minimum width=10mm, minimum height=4mm, fill=green!20, draw=green, thick}}
\tikzset{arrow/.style={line width=1mm}}
\tikzset{node distance = 0mm}
\coordinate (O) at (0,0); 
\coordinate (labelPosition) at (0,55mm); 
\draw[name path = sc1p, spacecraft] (0,0) circle [radius=\spacecraftRadius]
node (sc1) {};
\draw[name path = sc2p, spacecraft] (120:\spacecraftSeparation) circle [radius=\spacecraftRadius]
node (sc2) {};
\draw[name path = sc3p, spacecraft] (60:\spacecraftSeparation) circle [radius=\spacecraftRadius]
node (sc3) {};
\node (sc1l) at($(sc1.center)+(0,5mm)$) {\textbf{SC 1}};
\node (sc2l) at($(sc2.center)+(3mm,-1mm)$) {\textbf{SC 2}};
\node (sc3l) at($(sc3.center)+(-3mm,-1mm)$) {\textbf{SC 3}};
\coordinate (O1) at ($(sc1.center)+(0,-10mm)$); 
\coordinate (O2) at ($(sc2.center)+(150:10mm)$); 
\coordinate (O3) at ($(sc3.center)+(30:10mm)$); 
\path[name path = sc12] (O1) -- (O2);
\path[name path = sc13] (O1) -- (O3);
\path[name path = sc23] (O2) -- (O3);
\draw[name intersections={of=sc12 and sc1p, by=x}] node[masterlaser, rotate=-60] (la12) at(x) {\laserOneTwo};
\draw[name intersections={of=sc12 and sc2p, by=x}] node[laser, rotate=-60] (la21) at(x) {\laserTwoOne};
\draw[name intersections={of=sc13 and sc1p, by=x}] node[laser, rotate=+60] (la13) at(x) {\laserOneThree};
\draw[name intersections={of=sc13 and sc3p, by=x}] node[laser, rotate=+60] (la31) at(x) {\laserThreeOne};
\draw[name intersections={of=sc23 and sc2p, by=x}] node[laser, rotate=  0] (la23) at(x) {\laserTwoThree};
\draw[name intersections={of=sc23 and sc3p, by=x}] node[laser, rotate=  0] (la32) at(x) {\laserThreeTwo};
\draw[-latex, arrow, lred] (la12)--(la21); 
\draw[-latex, arrow, lblue] (la13)--(la31); 
\draw[-latex, arrow, lblue] (la12.east)..controls +(0.75,-1) and +(-0.75,-1)..(la13.west); 
\draw[-latex, arrow, lred] (la21.west)..controls +(-0.5,1.20) and +(-1.20,0.0)..(la23.west); 
\draw[-latex, arrow, lblue] (la31.east)..controls +(0.5,1.20) and +(1.20,0.0)..(la32.east); 
\node[] at (0, 2.5) {\large \textbf{N1-L12}};
\end{tikzpicture}
\end{minipage}
\hfill  
\begin{minipage}[t]{0.48\textwidth}  
\centering  
\begin{tikzpicture}
\definecolor{dkblue}{rgb} {0.00,0.33,0.68}
\definecolor{dkred}{rgb} {0.80,0.20,0.00}
\definecolor{lblue}{rgb} {0.00,0.60,0.95}
\definecolor{lred}{rgb} {0.95,0.30,0.00}
\tikzset{spacecraft/.style={circle, minimum width = 24mm, draw=black!60, fill=gray!10, very thick}}
\tikzset{laser/.style={rectangle, minimum width=10mm, minimum height=4mm, fill=red!20, draw=red!80, thick}}
\tikzset{masterlaser/.style={rectangle, minimum width=10mm, minimum height=4mm, fill=green!20, draw=green, thick}}
\tikzset{arrow/.style={line width=1mm}}
\tikzset{node distance = 0mm}
\coordinate (O) at (0,0); 
\coordinate (labelPosition) at (0,55mm); 
\draw[name path = sc1p, spacecraft] (0,0) circle [radius=\spacecraftRadius]
node (sc1) {};
\draw[name path = sc2p, spacecraft] (120:\spacecraftSeparation) circle [radius=\spacecraftRadius]
node (sc2) {};
\draw[name path = sc3p, spacecraft] (60:\spacecraftSeparation) circle [radius=\spacecraftRadius]
node (sc3) {};
\node (sc1l) at($(sc1.center)+(0,5mm)$) {\textbf{SC 1}};
\node (sc2l) at($(sc2.center)+(3mm,-1mm)$) {\textbf{SC 2}};
\node (sc3l) at($(sc3.center)+(-3mm,-1mm)$) {\textbf{SC 3}};
\coordinate (O1) at ($(sc1.center)+(0,-10mm)$); 
\coordinate (O2) at ($(sc2.center)+(150:10mm)$); 
\coordinate (O3) at ($(sc3.center)+(30:10mm)$); 
\path[name path = sc12] (O1) -- (O2);
\path[name path = sc13] (O1) -- (O3);
\path[name path = sc23] (O2) -- (O3);
\draw[name intersections={of=sc12 and sc1p, by=x}] node[masterlaser, rotate=-60] (la12) at(x) {\laserOneTwo};
\draw[name intersections={of=sc12 and sc2p, by=x}] node[laser, rotate=-60] (la21) at(x) {\laserTwoOne};
\draw[name intersections={of=sc13 and sc1p, by=x}] node[laser, rotate=+60] (la13) at(x) {\laserOneThree};
\draw[name intersections={of=sc13 and sc3p, by=x}] node[laser, rotate=+60] (la31) at(x) {\laserThreeOne};
\draw[name intersections={of=sc23 and sc2p, by=x}] node[laser, rotate=  0] (la23) at(x) {\laserTwoThree};
\draw[name intersections={of=sc23 and sc3p, by=x}] node[laser, rotate=  0] (la32) at(x) {\laserThreeTwo};
\draw[-latex, arrow, lblue] (la12.east)..controls +(0.75,-1) and +(-0.75,-1)..(la13.west); 
\draw[latex-, arrow, lblue] (la21.west)..controls +(-0.5,1.20) and +(-1.20,0.0)..(la23.west); 
\draw[-latex, arrow, lred]  (la12.west)..controls +(120:1)  and +(0,-0.5) ..(la23.south);  
\draw[-latex, arrow, lblue] (la13.east)..controls +(60:1)   and +(0,-0.5) ..(la32.south);  
\draw[-latex, arrow, lred]  (la23.east)..controls +(1,0)    and +(150:0.5) ..(la31.north); 
\node[] at (0, 2.5) {\large \textbf{S1-L12}};
\end{tikzpicture}
\end{minipage}
\caption{Graph representation of two possible laser locking schemes in LISA with laser ``LA12'' (colored in green) acting as the primary laser. The diagram on the left (N1-L12) shows a conventional (i.e., ``non-swap'') scheme, and the one on the right (S1-L12) shows a  frequency-swap scheme. The arrows indicate the existence of a transponder lock between the two linked lasers, where the laser pointed to by the arrow is the transponder. All six lasers in the constellation must be linked, directly or indirectly, to the primary laser. Note how in the non-swap case (N1-L12), the inter-spacecraft laser links are formed by connecting the two lasers pointed along the same arm (i.e., the local oscillator for each science interferometer link is also the transmit beam along the same arm), whereas in the frequency-swap case (S1-L12), the local oscillator is swapped to the other onboard laser.}
\label{figure:N1L12-S1L12}
\end{figure*}
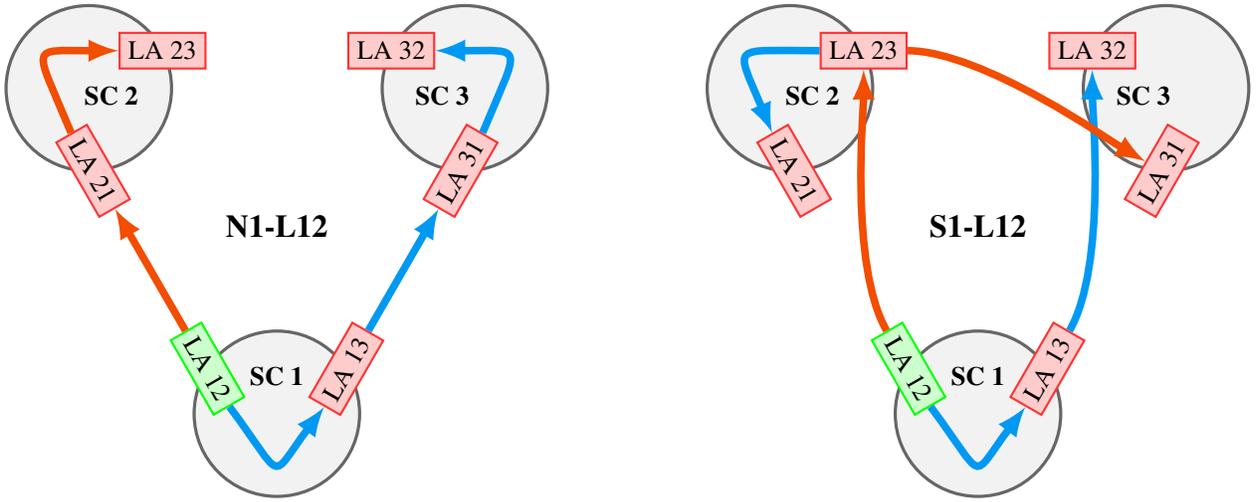

\subsection{Notation}

The notation follows our previous work~\cite{Heinzel2024}:
\begin{itemize}
\item $L_{12}$, $L_{21}$ $L_{13}$, $L_{31}$, $L_{23}$, $L_{32}$ refer to the 6 lasers in the constellation, more specifically to their \emph{laser frequencies}. The numbering $L_{ij}$ specifies that the laser is emitted from spacecraft $i$ and received by spacecraft $j$. 
\item $D_1$, $D_2$, $D_3$ refer to the three Doppler shifts that occur on each link, where $D_k$ denotes the link between the spacecrafts that do not include spacecraft $k$, i.e., $D_1$ refers to the Doppler shift affecting $L_{23}$ and $L_{32}$ and so on. We assume that they are the same in both directions, and moreover the same for each laser beam. Minor deviations due to the Sagnac effect and the differences in the laser frequencies are important for Time-Delay Interferometry (TDI) and data analysis, but not in our context.
\item $O_1$, $O_2$, $O_3$, $O_4$, $O_5$ are the five offset frequencies that are used to phase-lock each transponder laser to some other laser frequency which is eventually derived from the primary laser. 
\item $B_{11}$, $B_{12}$, \dots{} $B_{33}$ are a total of nine beatnote frequencies that occur in the constellation.  They are listed in detail below.
\end{itemize}
The laser frequencies $L_{ij}$ are always assumed to be positive, about 280\,THz, although that number never plays a direct role. The quantities $D_i$, $O_i$ and $B_{ij}$, however, arise as differences between laser frequencies and can have both signs.

To demonstrate the generality of our method, we apply it to both the swap and non-swap configurations. Frequency swapping applies to either all or none of the lasers, since it implies different ``hardwired'' routing of laser beams on the optical bench. See Figure~\ref{figure:N1L12-S1L12} for a graph representation of the first locking configuration in each scheme with laser L12 chosen as primary. 

In mathematical terms, frequency swapping modifies the definition of the science interferometer beatnotes. In the non-swap scheme, for spacecraft $i$, these beatnotes are given by:
\begin{alignat}{1}
B_{ij} &= L_{ji} + D_{k} - L_{ij}, \\
B_{ik} &= L_{ki} + D_{j} - L_{ik}.
\end{alignat}
In the frequency-swap scheme, the corresponding beatnotes are:
\begin{alignat}{1}
B_{ij} &= L_{ji} + D_{k} - L_{ik}, \\
B_{ik} &= L_{ki} + D_{j} - L_{ij}.
\end{alignat}
The first two terms on the right-hand side of these equations represent the frequency of the received beam, which has accumulated a Doppler shift. The last term corresponds to the local oscillator frequency. In the frequency-swap scheme, the role of the local oscillator laser is exchanged between the two onboard lasers: the science interferometer beatnote $(ij)$ is formed using the laser transmitted to spacecraft $k$, rather than the one transmitted to spacecraft $j$.


\subsection{Exhaustive Search of Locking Configurations}

The method identifies all feasible laser locking schemes by first enumerating all possible configurations and subsequently evaluating their validity. As an illustrative example, we consider the valid non-swap configuration labeled $N3-L32$, see Figure~\ref{figure:N3-L32}.  The five steps outlined below were independently carried out using two distinct symbolic computation frameworks, both yielding identical results.

\begin{description}

\item[Step 1 - Choice of primary laser] We iterate over all possible choices of primary laser. In LISA, any of the six lasers in the constellation can be locked to an ultra-stable frequency reference and act as the primary laser.

\item[Step 2 - Enumeration of all possible transponder schemes] Given a choice of primary laser, the remaining five lasers are \emph{secondary} or \emph{transponders}. We iterate over all possible permutations of transponder locks. The secondary laser $L_{ij}$ can be locked to either the other onboard laser in the local spacecraft ($L_{ik}$), or to the light received from one of the lasers onboard the remote spacecraft $j$ ($L_{jk}$, where $k=i$ in the non-swap configuration). This yields $2^5 = 32$ potential locking scheme candidates for a given primary laser, or $6 \cdot 2^5 = 192$ total candidates taking into account permutations of the primary laser. However, only a subset of these are useful—those where all secondary lasers are directly or indirectly referenced exclusively to the primary laser. This constraint is enforced in the subsequent steps.

\item[Step 3 - Setup of a system of linear equations] Each of the configurations enumerated in Step 2 is described by a set of linear relations among the laser frequencies, with the introduction of the transponder offset frequencies $O_1...O_5$. For example, for $N3-L32$:
\begin{align}
& L_{32} \text{ primary},\nonumber \\
& L_{23} = L_{32} + O_1,\nonumber \\
& L_{13} = L_{12} + O_2,\nonumber \\
& L_{31} = L_{32} + O_3,\nonumber \\
& L_{21} = L_{23} + O_4,\nonumber \\
& L_{12} = L_{21} + O_5.
\end{align}


\item[Step 4 - Insertion of Doppler shifts]
The appropriate frequency shifts due to the Doppler effect are inserted in the locks involving an inter-satellite link. In this example:
\begin{align}
& L_{32} \text{ primary}\nonumber \\
& L_{23} = L_{32}  + O_1 + D_1,\nonumber \\
& L_{13}  = L_{12} + O_2,\nonumber \\
& L_{31} = L_{32}  + O_3,\nonumber \\
& L_{21}  = L_{23} + O_4,\nonumber \\
& L_{12} = L_{21}  + O_5 + D_3.
\label{eq:initial}
\end{align}

\item[Step 5 - Substitution of laser frequencies]
Next, all laser frequencies on the right hand sides other than the primary laser are iteratively replaced by their corresponding expressions from Equation~(\ref{eq:initial}). This is repeated until the equations do not change any more. By then, only the primary laser should appear on the right hand sides, exactly once for each transponder laser. Otherwise, the scheme is invalid and discarded. In the example:
\begin{align}
& L_{32} \text{ (primary)} \nonumber \\
& L_{23} = L_{32} + D_1 + O_1, \nonumber \\
& L_{13} = L_{32} +D_1 + D_3 + O_1 + O_2 + O_4 + O_5, \nonumber \\
& L_{31} = L_{32} + O_3, \nonumber \\
& L_{21} = L_{32} +D_1 + O_1 + O_4, \nonumber \\
& L_{12} = L_{32} +D_1 + D_3 + O_1 + O_4 + O_5.
\label{eq:final}
\end{align}

Performing this exhaustive analysis reveals that out of the 32 total candidates for a fixed primary laser, there are exactly:
\begin{itemize}
\item 6 valid and unique configurations in the case of non-swap, and
\item 12 valid and unique configurations for frequency swap.
\end{itemize}
Considering all permutations of the primary laser, out of the 192 total candidates, there remain $6 \cdot 6 = 36$ valid configurations in the case of non-swap, and $6 \cdot 12 = 72$ valid configurations for frequency swap.
\end{description}

\subsection{Construction of the beatnotes}

The principal quantities that the frequency planning has to monitor are the frequencies of the nine beatnotes, which are constructed next. Recall that $B_{ij}$ is the beatnote that occurs in spacecraft $i$ and the second interfering laser comes from spacecraft $j$. 
In particular, for non-swap we have:
\begin{alignat}{2}
B_{11} &= L_{13} - L_{12},        \quad && \text{on SC 1}\nonumber \\
B_{12} &= L_{21} + D_3 - L_{12},   \quad && \text{on SC 1}\nonumber \\
B_{13} &= L_{31} + D_2 - L_{13},   \quad && \text{on SC 1}\nonumber \\
B_{21} &= L_{12} + D_3 - L_{21},   \quad && \text{on SC 2}\nonumber \\
B_{22} &= L_{21} - L_{23},        \quad && \text{on SC 2}\nonumber \\
B_{23} &= L_{32} + D_1 - L_{23},   \quad && \text{on SC 2}\nonumber \\
B_{31} &= L_{13} + D_2 - L_{31},   \quad && \text{on SC 3}\nonumber \\
B_{32} &= L_{23} + D_1 - L_{32},   \quad && \text{on SC 3}\nonumber \\
B_{33} &= L_{32} - L_{31},        \quad && \text{on SC 3},
\label{eq:beatn}
\end{alignat}
whereas for frequency swap we get
\begin{alignat}{2}
B_{11} &= L_{13} - L_{12}          \quad && \text{on SC1}\nonumber \\
B_{12} &= L_{21} + D_3 - L_{13}    \quad && \text{on SC1}\nonumber \\
B_{13} &= L_{31} + D_2 - L_{12}    \quad && \text{on SC1}\nonumber \\
B_{21} &= L_{12} + D_3 - L_{23}   \quad && \text{on SC2}\nonumber \\
B_{22} &= L_{21} - L_{23}         \quad && \text{on SC2}\nonumber \\
B_{23} &= L_{32} + D_1 - L_{21}   \quad && \text{on SC2}\nonumber \\
B_{31} &= L_{13} + D_2 - L_{32}   \quad && \text{on SC3}\nonumber \\
B_{32} &= L_{23} + D_1 - L_{31}   \quad && \text{on SC3}\nonumber \\
B_{33} &= L_{32} - L_{31}         \quad && \text{on SC3}
\label{eq:beats}
\end{alignat}
Note that Equations~(\ref{eq:beatn}) and   (\ref{eq:beats}) are generally valid for all configurations, whereas
 Equations~(\ref{eq:initial}), (\ref{eq:final}) as well as (\ref{eq:beatn2}) below use one particular configuration, $N3-L32$, as  illustrative example.

The expressions for the laser frequencies from the previous step are now inserted into these equations, resulting for this particular configuration in:
\begin{alignat}{2}
B_{11}  &=  O_2,                                                \nonumber \\
B_{12}  &=  - O_5,                                           \nonumber \\
B_{13}  &=  - D_1 + D_2 - D_3 - O_1 - O_2 + O_3 - O_4 - O_5,        \nonumber \\
B_{21}  &=  2 D_3 + O_5,                                    \nonumber \\
B_{22}  &=  O_4,                                                \nonumber \\
B_{23}  &=  - O_1,                                           \nonumber \\
B_{31}  &=  D_1 + D_2 + D_3 + O_1 + O_2 - O_3 + O_4 + O_5,        \nonumber \\
B_{32}  &=  2 D_1 + O_1,                                    \nonumber \\
B_{33}  &=  - O_3.                                                
\label{eq:beatn2}
\end{alignat}


\subsection{Beatnote classification}

We propose a beatnote classification based on whether or not a certain beatnote is used to lock a transponder laser. There are five of those ($B_{11}$, $B_{12}$, $B_{22}$, $B_{23}$ and $B_{33}$ in the example discussed above), which will be called ``locking'' beatnotes, and four that are not used in laser locks, thus called ``non-locking'' beatnotes. Apart from receiving a different treatment in the frequency plan optimization, there is also an operational distinction: If any of the five locking beatnote frequencies passes through a forbidden frequency range, the respective offset laser lock will temporarily be lost. Depending on where in the locking chain that beatnote is located, the loss of lock might ripple through to other locks. A lost lock requires initiating a re-acquisition procedure. This is not a disaster as such, since such a procedure must be available in any case, but it would be preferable if it can be avoided.

A transition through the forbidden region of one of the non-locking beatnotes, on the other hand, will not cause any transponder laser phase lock to drop out of lock. Any transition of any beatnote through the forbidden region will, however, cause an interruption in the science data. Such an interruption may be short (order of seconds to minutes) and not significantly affecting LISA science. Nevertheless, it is one aim of frequency planning to minimize the number of such interruptions.

According to the above discussion, the beatnotes are re-arranged into locking and non-locking ones, and reported as follows for our example:

\begin{align}
\begin{pmatrix}
B_{23}\\
B_{11}\\
B_{33}\\
B_{22}\\
B_{12}\\
B_{13}\\
B_{21}\\
B_{31}\\
B_{32}
\end{pmatrix}
=         
\begin{pmatrix}
       0 & 0 & 0  &  -1  &   0 &    0   &  0 &    0    \\   
       0 & 0 & 0  &   0  &   1 &    0   &  0 &    0    \\
       0 & 0 & 0  &   0  &   0 &   -1   &  0 &    0    \\
       0 & 0 & 0  &   0  &   0 &    0   &  1 &    0    \\
       0 & 0 & 0  &   0  &   0 &    0   &  0 &   -1    \\
      -1 & 1 &-1  &  -1  &  -1 &    1   & -1 &   -1    \\
       0 & 0 & 2  &   0  &   0 &    0   &  0 &    1    \\
       1 & 1 & 1  &   1  &   1 &   -1   &  1 &    1    \\
       2 & 0 & 0  &   1  &   0 &    0   &  0 &    0    
\end{pmatrix}
 \cdot
\begin{pmatrix}
D_1 \\
D_2 \\
D_3 \\
O_1 \\
O_2 \\
O_3 \\
O_4 \\
O_5 
\end{pmatrix}
\label{eq:mat}
\end{align}

The different blocks in this matrix play a crucial role in the frequency planning optimization method, since they determine linear inequality constraints for the offsets upon imposing physical constraints on the beatnotes, (such as that they stay within a certain frequency band, and that there is no crossing in value with respect to the local beatnote on each spacecraft.)

With this method at hand we generate all possible valid locking schemes for any possible primary laser, obtaining the laser locking configuration relationships analogous to the system of equations \ref{eq:final}, and the beatnote dependencies such as \ref{eq:mat}.

\renewcommand{\spacecraftSeparation}{38mm}
\renewcommand{\spacecraftRadius}{10.5mm}

\begin{figure*}
\centering
\begin{subfigure}{\textwidth}
\centering
\begin{minipage}{0.3\textwidth}
\centering
\begin{tikzpicture}[scale=1.0]
\definecolor{dkblue}{rgb} {0.00,0.33,0.68}
\definecolor{dkred}{rgb} {0.80,0.20,0.00}
\definecolor{lblue}{rgb} {0.00,0.60,0.95}
\definecolor{lred}{rgb} {0.95,0.30,0.00}
\tikzset{spacecraft/.style={circle, minimum width = 24mm, draw=black!60, fill=gray!10, very thick}}
\tikzset{laser/.style={rectangle, minimum width=10mm, minimum height=4mm, fill=red!20, draw=red!80, thick}}
\tikzset{masterlaser/.style={rectangle, minimum width=10mm, minimum height=4mm, fill=green!20, draw=green, thick}}
\tikzset{arrow/.style={line width=1mm}}
\tikzset{node distance = 0mm}
\coordinate (O) at (0,0); 
\coordinate (labelPosition) at (0,55mm); 
\draw[name path = sc1p, spacecraft] (0,0) circle [radius=\spacecraftRadius]
node (sc1) {};
\draw[name path = sc2p, spacecraft] (120:\spacecraftSeparation) circle [radius=\spacecraftRadius]
node (sc2) {};
\draw[name path = sc3p, spacecraft] (60:\spacecraftSeparation) circle [radius=\spacecraftRadius]
node (sc3) {};
\node (sc1l) at($(sc1.center)+(0,5mm)$) {\textbf{SC 1}};
\node (sc2l) at($(sc2.center)+(3mm,-1mm)$) {\textbf{SC 2}};
\node (sc3l) at($(sc3.center)+(-3mm,-1mm)$) {\textbf{SC 3}};
\coordinate (O1) at ($(sc1.center)+(0,-10mm)$); 
\coordinate (O2) at ($(sc2.center)+(150:10mm)$); 
\coordinate (O3) at ($(sc3.center)+(30:10mm)$); 
\path[name path = sc12] (O1) -- (O2);
\path[name path = sc13] (O1) -- (O3);
\path[name path = sc23] (O2) -- (O3);
\draw[name intersections={of=sc12 and sc1p, by=x}] node[laser, rotate=-60] (la12) at(x) {\laserOneTwo};
\draw[name intersections={of=sc12 and sc2p, by=x}] node[laser, rotate=-60] (la21) at(x) {\laserTwoOne};
\draw[name intersections={of=sc13 and sc1p, by=x}] node[laser, rotate=+60] (la13) at(x) {\laserOneThree};
\draw[name intersections={of=sc13 and sc3p, by=x}] node[laser, rotate=+60] (la31) at(x) {\laserThreeOne};
\draw[name intersections={of=sc23 and sc2p, by=x}] node[laser, rotate=  0] (la23) at(x) {\laserTwoThree};
\draw[name intersections={of=sc23 and sc3p, by=x}] node[masterlaser, rotate=  0] (la32) at(x) {\laserThreeTwo};
\draw[latex-, arrow, lblue] (la23)--(la32); 
\draw[latex-, arrow, lred] (la13)--(la31); 
\draw[latex-, arrow, lred] (la12.east)..controls +(0.75,-1) and +(-0.75,-1)..(la13.west); 
\draw[latex-, arrow, lblue] (la21.west)..controls +(-0.5,1.20) and +(-1.20,0.0)..(la23.west); 
\draw[latex-, arrow, lred] (la31.east)..controls +(0.5,1.20) and +(1.20,0.0)..(la32.east); 
\node[] at (0, 2) {\large \textbf{N1-L32}};
\end{tikzpicture}
\end{minipage}
\hfill
\begin{minipage}{0.3\textwidth}
\centering
\begin{alignat*}{3}
& L_{32} &&= L_{32} \\
& L_{23} &&= L_{32} + O_{1} + D_{1} \\
& L_{13} &&= L_{31} + O_{2} + D_{2} \\
& L_{31} &&= L_{32} + O_{3} \\
& L_{21} &&= L_{23} + O_{4} \\
& L_{12} &&= L_{13} + O_{5} \\
\midrule
& L_{32} &&= L_{32} \\
& L_{23} &&= D_{1} + L_{32} + O_{1} \\
& L_{13} &&= D_{2} + L_{32} + O_{2} + O_{3} \\
& L_{31} &&= L_{32} + O_{3} \\
& L_{21} &&= D_{1} + L_{32} + O_{1} + O_{4} \\
& L_{12} &&= D_{2} + L_{32} + O_{2} + O_{3} + O_{5}
\end{alignat*}
\end{minipage}
\begin{minipage}{0.3\textwidth}
\begin{center}
\vspace{5mm}
\renewcommand{\arraystretch}{1.5}
\begin{tabular}{ c | c c c c c c c c }
~ & $D_{1}$ & $D_{2}$ & $D_{3}$ & $O_{1}$ & $O_{2}$ & $O_{3}$ & $O_{4}$ & $O_{5}$ \\
\hline
$B_{23}$ & 0 & 0 & 0 & -1 & 0 & 0 & 0 & 0 \\
$B_{13}$ & 0 & 0 & 0 & 0 & -1 & 0 & 0 & 0 \\
$B_{33}$ & 0 & 0 & 0 & 0 & 0 & -1 & 0 & 0 \\
$B_{22}$ & 0 & 0 & 0 & 0 & 0 & 0 & 1 & 0 \\
$B_{11}$ & 0 & 0 & 0 & 0 & 0 & 0 & 0 & -1 \\
$B_{12}$ & 1 & -1 & 1 & 1 & -1 & -1 & 1 & -1 \\
$B_{21}$ & -1 & 1 & 1 & -1 & 1 & 1 & -1 & 1 \\
$B_{31}$ & 0 & 2 & 0 & 0 & 1 & 0 & 0 & 0 \\
$B_{32}$ & 2 & 0 & 0 & 1 & 0 & 0 & 0 & 0 \\
\end{tabular}
\end{center}
\end{minipage}
\caption{Configuration N1-L32, with three local locks and one inter-spacecraft lock.}
\label{figure:N1-L32}
\end{subfigure}
\begin{subfigure}{\textwidth}
\begin{minipage}{0.3\textwidth}
\begin{tikzpicture}
\definecolor{dkblue}{rgb} {0.00,0.33,0.68}
\definecolor{dkred}{rgb} {0.80,0.20,0.00}
\definecolor{lblue}{rgb} {0.00,0.60,0.95}
\definecolor{lred}{rgb} {0.95,0.30,0.00}
\tikzset{spacecraft/.style={circle, minimum width = 24mm, draw=black!60, fill=gray!10, very thick}}
\tikzset{laser/.style={rectangle, minimum width=10mm, minimum height=4mm, fill=red!20, draw=red!80, thick}}
\tikzset{masterlaser/.style={rectangle, minimum width=10mm, minimum height=4mm, fill=green!20, draw=green, thick}}
\tikzset{arrow/.style={line width=1mm}}
\tikzset{node distance = 0mm}
\coordinate (O) at (0,0); 
\coordinate (labelPosition) at (0,55mm); 
\draw[name path = sc1p, spacecraft] (0,0) circle [radius=\spacecraftRadius]
node (sc1) {};
\draw[name path = sc2p, spacecraft] (120:\spacecraftSeparation) circle [radius=\spacecraftRadius]
node (sc2) {};
\draw[name path = sc3p, spacecraft] (60:\spacecraftSeparation) circle [radius=\spacecraftRadius]
node (sc3) {};
\node (sc1l) at($(sc1.center)+(0,5mm)$) {\textbf{SC 1}};
\node (sc2l) at($(sc2.center)+(3mm,-1mm)$) {\textbf{SC 2}};
\node (sc3l) at($(sc3.center)+(-3mm,-1mm)$) {\textbf{SC 3}};
\coordinate (O1) at ($(sc1.center)+(0,-10mm)$); 
\coordinate (O2) at ($(sc2.center)+(150:10mm)$); 
\coordinate (O3) at ($(sc3.center)+(30:10mm)$); 
\path[name path = sc12] (O1) -- (O2);
\path[name path = sc13] (O1) -- (O3);
\path[name path = sc23] (O2) -- (O3);
\draw[name intersections={of=sc12 and sc1p, by=x}] node[laser, rotate=-60] (la12) at(x) {\laserOneTwo};
\draw[name intersections={of=sc12 and sc2p, by=x}] node[laser, rotate=-60] (la21) at(x) {\laserTwoOne};
\draw[name intersections={of=sc13 and sc1p, by=x}] node[laser, rotate=+60] (la13) at(x) {\laserOneThree};
\draw[name intersections={of=sc13 and sc3p, by=x}] node[laser, rotate=+60] (la31) at(x) {\laserThreeOne};
\draw[name intersections={of=sc23 and sc2p, by=x}] node[laser, rotate=  0] (la23) at(x) {\laserTwoThree};
\draw[name intersections={of=sc23 and sc3p, by=x}] node[masterlaser, rotate=  0] (la32) at(x) {\laserThreeTwo};
\draw[latex-, arrow, lblue] (la12)--(la21); 
\draw[latex-, arrow, lblue] (la23)--(la32); 
\draw[latex-, arrow, lred] (la13)--(la31); 
\draw[latex-, arrow, lblue] (la21.west)..controls +(-0.5,1.20) and +(-1.20,0.0)..(la23.west); 
\draw[latex-, arrow, lred] (la31.east)..controls +(0.5,1.20) and +(1.20,0.0)..(la32.east); 
\node[] at (0, 2) {\large \textbf{N2-L32}};
\end{tikzpicture}
\end{minipage}
\hfill
\begin{minipage}{0.3\textwidth}
\begin{alignat*}{3}
& L_{32} &&= L_{32} \\
&L_{23} &&= L_{32} + O_{1} + D_{1} \\
&L_{13} &&= L_{31} + O_{2} + D_{2} \\
&L_{31} &&= L_{32} + O_{3} \\
&L_{21} &&= L_{23} + O_{4} \\
&L_{12} &&= L_{21} + O_{5} + D_{3} \\
\midrule
& L_{32} &&= L_{32} \\
&L_{23} &&= D_{1} + L_{32} + O_{1} \\
&L_{13} &&= D_{2} + L_{32} + O_{2} + O_{3} \\
&L_{31} &&= L_{32} + O_{3} \\
&L_{21} &&= D_{1} + L_{32} + O_{1} + O_{4} \\
&L_{12} &&= D_{1} + D_{3} + L_{32} + O_{1} + O_{4} + O_{5} 
\end{alignat*}
\end{minipage}
\begin{minipage}{0.3\textwidth}
\begin{center}
\vspace{5mm}
\renewcommand{\arraystretch}{1.5}
\begin{tabular}{c | c c c c c c c c }
~ & $D_{1}$ & $D_{2}$ & $D_{3}$ & $O_{1}$ & $O_{2}$ & $O_{3}$ & $O_{4}$ & $O_{5}$ \\
\hline
$B_{23}$ & 0 & 0 & 0 & -1 & 0 & 0 & 0 & 0 \\
$B_{13}$ & 0 & 0 & 0 & 0 & -1 & 0 & 0 & 0 \\
$B_{33}$ & 0 & 0 & 0 & 0 & 0 & -1 & 0 & 0 \\
$B_{22}$ & 0 & 0 & 0 & 0 & 0 & 0 & 1 & 0 \\
$B_{12}$ & 0 & 0 & 0 & 0 & 0 & 0 & 0 & -1 \\
$B_{11}$ & -1 & 1 & -1 & -1 & 1 & 1 & -1 & -1 \\
$B_{21}$ & 0 & 0 & 2 & 0 & 0 & 0 & 0 & 1 \\
$B_{31}$ & 0 & 2 & 0 & 0 & 1 & 0 & 0 & 0 \\
$B_{32}$ & 2 & 0 & 0 & 1 & 0 & 0 & 0 & 0 \\
\end{tabular}
\end{center}
\end{minipage}
\caption{Configuration N2-L32, with two local locks and two inter-spacecraft locks}
\label{figure:N2-L32}
\end{subfigure}
\begin{subfigure}{\textwidth}\label{subfig:N3-L32}
\begin{minipage}{0.3\textwidth}
\begin{tikzpicture}
\definecolor{dkblue}{rgb} {0.00,0.33,0.68}
\definecolor{dkred}{rgb} {0.80,0.20,0.00}
\definecolor{lblue}{rgb} {0.00,0.60,0.95}
\definecolor{lred}{rgb} {0.95,0.30,0.00}
\tikzset{spacecraft/.style={circle, minimum width = 24mm, draw=black!60, fill=gray!10, very thick}}
\tikzset{laser/.style={rectangle, minimum width=10mm, minimum height=4mm, fill=red!20, draw=red!80, thick}}
\tikzset{masterlaser/.style={rectangle, minimum width=10mm, minimum height=4mm, fill=green!20, draw=green, thick}}
\tikzset{arrow/.style={line width=1mm}}
\tikzset{node distance = 0mm}
\coordinate (O) at (0,0); 
\coordinate (labelPosition) at (0,55mm); 
\draw[name path = sc1p, spacecraft] (0,0) circle [radius=\spacecraftRadius]
node (sc1) {};
\draw[name path = sc2p, spacecraft] (120:\spacecraftSeparation) circle [radius=\spacecraftRadius]
node (sc2) {};
\draw[name path = sc3p, spacecraft] (60:\spacecraftSeparation) circle [radius=\spacecraftRadius]
node (sc3) {};
\node (sc1l) at($(sc1.center)+(0,5mm)$) {\textbf{SC 1}};
\node (sc2l) at($(sc2.center)+(3mm,-1mm)$) {\textbf{SC 2}};
\node (sc3l) at($(sc3.center)+(-3mm,-1mm)$) {\textbf{SC 3}};
\coordinate (O1) at ($(sc1.center)+(0,-10mm)$); 
\coordinate (O2) at ($(sc2.center)+(150:10mm)$); 
\coordinate (O3) at ($(sc3.center)+(30:10mm)$); 
\path[name path = sc12] (O1) -- (O2);
\path[name path = sc13] (O1) -- (O3);
\path[name path = sc23] (O2) -- (O3);
\draw[name intersections={of=sc12 and sc1p, by=x}] node[laser, rotate=-60] (la12) at(x) {\laserOneTwo};
\draw[name intersections={of=sc12 and sc2p, by=x}] node[laser, rotate=-60] (la21) at(x) {\laserTwoOne};
\draw[name intersections={of=sc13 and sc1p, by=x}] node[laser, rotate=+60] (la13) at(x) {\laserOneThree};
\draw[name intersections={of=sc13 and sc3p, by=x}] node[laser, rotate=+60] (la31) at(x) {\laserThreeOne};
\draw[name intersections={of=sc23 and sc2p, by=x}] node[laser, rotate=  0] (la23) at(x) {\laserTwoThree};
\draw[name intersections={of=sc23 and sc3p, by=x}] node[masterlaser, rotate=  0] (la32) at(x) {\laserThreeTwo};
\draw[latex-, arrow, lblue] (la12)--(la21); 
\draw[latex-, arrow, lblue] (la23)--(la32); 
\draw[-latex, arrow, lblue] (la12.east)..controls +(0.75,-1) and +(-0.75,-1)..(la13.west); 
\draw[latex-, arrow, lblue] (la21.west)..controls +(-0.5,1.20) and +(-1.20,0.0)..(la23.west); 
\draw[latex-, arrow, lred] (la31.east)..controls +(0.5,1.20) and +(1.20,0.0)..(la32.east); 
\node[] at (0, 2) {\large \textbf{N3-L32}};
\end{tikzpicture}
\end{minipage}
\hfill
\begin{minipage}{0.3\textwidth}
\hspace{-1cm}
\begin{alignat*}{3}
& L_{32} &&= L_{32} \\
&L_{23} &&= L_{32} + O_{1} + D_{1} \\
&L_{13} &&= L_{12} + O_{2} \\
&L_{31} &&= L_{32} + O_{3} \\
&L_{21} &&= L_{23} + O_{4} \\
&L_{12} &&= L_{21} + O_{5} + D_{3} \\
\midrule
& L_{32} &&= L_{32} \\
&L_{23} &&= D_{1} + L_{32} + O_{1} \\
&L_{13} &&= D_{1} + D_{3} + L_{32} + O_{1} + O_{2} + O_{4} + O_{5} \\
&L_{31} &&= L_{32} + O_{3} \\
&L_{21} &&= D_{1} + L_{32} + O_{1} + O_{4} \\
&L_{12} &&= D_{1} + D_{3} + L_{32} + O_{1} + O_{4} + O_{5} 
\end{alignat*}
\end{minipage}
\begin{minipage}{0.3\textwidth}
\begin{center}
\vspace{5mm}
\renewcommand{\arraystretch}{1.5}
\begin{tabular}{c | c c c c c c c c }
~ & $D_{1}$ & $D_{2}$ & $D_{3}$ & $O_{1}$ & $O_{2}$ & $O_{3}$ & $O_{4}$ & $O_{5}$ \\
\hline
$B_{23}$ & 0 & 0 & 0 & -1 & 0 & 0 & 0 & 0 \\
$B_{11}$ & 0 & 0 & 0 & 0 & 1 & 0 & 0 & 0 \\
$B_{33}$ & 0 & 0 & 0 & 0 & 0 & -1 & 0 & 0 \\
$B_{22}$ & 0 & 0 & 0 & 0 & 0 & 0 & 1 & 0 \\
$B_{12}$ & 0 & 0 & 0 & 0 & 0 & 0 & 0 & -1 \\
$B_{13}$ & -1 & 1 & -1 & -1 & -1 & 1 & -1 & -1 \\
$B_{21}$ & 0 & 0 & 2 & 0 & 0 & 0 & 0 & 1  \\
$B_{31}$ & 1 & 1 & 1 & 1 & 1 & -1 & 1 & 1  \\
$B_{32}$ & 2 & 0 & 0 & 1 & 0 & 0 & 0 & 0  \\
\end{tabular}
\end{center}
\end{minipage}
\caption{Configuration N3-L32, with three local locks and one inter-spacecraft lock.}
\label{figure:N3-L32}
\end{subfigure}
\caption{Laser locking configurations N1-L32, N2-L32, and N3-L32, with LA 32 as the primary laser.}
\label{figure:N1-N3-L32}
\end{figure*}

\begin{figure*}
\centering
\begin{subfigure}{\textwidth}
\begin{minipage}{0.3\textwidth}
\centering
\begin{tikzpicture}
\definecolor{dkblue}{rgb} {0.00,0.33,0.68}
\definecolor{dkred}{rgb} {0.80,0.20,0.00}
\definecolor{lblue}{rgb} {0.00,0.60,0.95}
\definecolor{lred}{rgb} {0.95,0.30,0.00}
\tikzset{spacecraft/.style={circle, minimum width = 24mm, draw=black!60, fill=gray!10, very thick}}
\tikzset{laser/.style={rectangle, minimum width=10mm, minimum height=4mm, fill=red!20, draw=red!80, thick}}
\tikzset{masterlaser/.style={rectangle, minimum width=10mm, minimum height=4mm, fill=green!20, draw=green, thick}}
\tikzset{arrow/.style={line width=1mm}}
\tikzset{node distance = 0mm}
\coordinate (O) at (0,0); 
\coordinate (labelPosition) at (0,55mm); 
\draw[name path = sc1p, spacecraft] (0,0) circle [radius=\spacecraftRadius]
node (sc1) {};
\draw[name path = sc2p, spacecraft] (120:\spacecraftSeparation) circle [radius=\spacecraftRadius]
node (sc2) {};
\draw[name path = sc3p, spacecraft] (60:\spacecraftSeparation) circle [radius=\spacecraftRadius]
node (sc3) {};
\node (sc1l) at($(sc1.center)+(0,5mm)$) {\textbf{SC 1}};
\node (sc2l) at($(sc2.center)+(3mm,-1mm)$) {\textbf{SC 2}};
\node (sc3l) at($(sc3.center)+(-3mm,-1mm)$) {\textbf{SC 3}};
\coordinate (O1) at ($(sc1.center)+(0,-10mm)$); 
\coordinate (O2) at ($(sc2.center)+(150:10mm)$); 
\coordinate (O3) at ($(sc3.center)+(30:10mm)$); 
\path[name path = sc12] (O1) -- (O2);
\path[name path = sc13] (O1) -- (O3);
\path[name path = sc23] (O2) -- (O3);
\draw[name intersections={of=sc12 and sc1p, by=x}] node[laser, rotate=-60] (la12) at(x) {\laserOneTwo};
\draw[name intersections={of=sc12 and sc2p, by=x}] node[laser, rotate=-60] (la21) at(x) {\laserTwoOne};
\draw[name intersections={of=sc13 and sc1p, by=x}] node[laser, rotate=+60] (la13) at(x) {\laserOneThree};
\draw[name intersections={of=sc13 and sc3p, by=x}] node[laser, rotate=+60] (la31) at(x) {\laserThreeOne};
\draw[name intersections={of=sc23 and sc2p, by=x}] node[laser, rotate=  0] (la23) at(x) {\laserTwoThree};
\draw[name intersections={of=sc23 and sc3p, by=x}] node[masterlaser, rotate=  0] (la32) at(x) {\laserThreeTwo};
\draw[-latex, arrow, lred] (la12)--(la21); 
\draw[latex-, arrow, lblue] (la23)--(la32); 
\draw[latex-, arrow, lred] (la13)--(la31); 
\draw[latex-, arrow, lred] (la12.east)..controls +(0.75,-1) and +(-0.75,-1)..(la13.west); 
\draw[latex-, arrow, lred] (la31.east)..controls +(0.5,1.20) and +(1.20,0.0)..(la32.east); 
\node[] at (0, 2) {\large \textbf{N4-L32}};
\end{tikzpicture}
\end{minipage}
\hfill
\begin{minipage}{0.3\textwidth}
\begin{alignat*}{3}
&L_{32} &&= L_{32} \\
&L_{23} &&= L_{32} + O_{1} + D_{1} \\
&L_{13} &&= L_{31} + O_{2} + D_{2} \\
&L_{31} &&= L_{32} + O_{3} \\
&L_{21} &&= L_{12} + O_{4} + D_{3} \\
&L_{12} &&= L_{13} + O_{5} \\
\midrule
& L_{32} &&= L_{32} \\
&L_{23} &&= D_{1} + L_{32} + O_{1} \\
&L_{13} &&= D_{2} + L_{32} + O_{2} + O_{3} \\
&L_{31} &&= L_{32} + O_{3} \\
&L_{21} &&= D_{2} + D_{3} + L_{32} + \Sigma_{i=2}^{5} O_{i} \\
&L_{12} &&= D_{2} + L_{32} + O_{2} + O_{3} + O_{5} 
\end{alignat*}
\end{minipage}
\begin{minipage}{0.3\textwidth}
\begin{center}
\vspace{5mm}
\renewcommand{\arraystretch}{1.5}
\begin{tabular}{c | c c c c c c c c}
~ & $D_{1}$ & $D_{2}$ & $D_{3}$ & $O_{1}$ & $O_{2}$ & $O_{3}$ & $O_{4}$ & $O_{5}$ \\
\hline
$B_{23}$ & 0 & 0 & 0 & -1 & 0 & 0 & 0 & 0 \\
$B_{13}$ & 0 & 0 & 0 & 0 & -1 & 0 & 0 & 0 \\
$B_{33}$ & 0 & 0 & 0 & 0 & 0 & -1 & 0 & 0 \\
$B_{21}$ & 0 & 0 & 0 & 0 & 0 & 0 & -1 & 0 \\
$B_{11}$ & 0 & 0 & 0 & 0 & 0 & 0 & 0 & -1 \\
$B_{12}$ & 0 & 0 & 2 & 0 & 0 & 0 & 1 & 0 \\
$B_{22}$ & -1 & 1 & 1 & -1 & 1 & 1 & 1 & 1 \\
$B_{31}$ & 0 & 2 & 0 & 0 & 1 & 0 & 0 & 0 \\
$B_{32}$ & 2 & 0 & 0 & 1 & 0 & 0 & 0 & 0 \\
\end{tabular}
\end{center}
\end{minipage}
\caption{Configuration N4-L32, with two local locks and two inter-spacecraft locks.}
\label{figure:N4-L32}
\end{subfigure}
\begin{subfigure}{\textwidth}
\centering
\begin{minipage}{0.3\textwidth}
\centering
\begin{tikzpicture}
\definecolor{dkblue}{rgb} {0.00,0.33,0.68}
\definecolor{dkred}{rgb} {0.80,0.20,0.00}
\definecolor{lblue}{rgb} {0.00,0.60,0.95}
\definecolor{lred}{rgb} {0.95,0.30,0.00}
\tikzset{spacecraft/.style={circle, minimum width = 24mm, draw=black!60, fill=gray!10, very thick}}
\tikzset{laser/.style={rectangle, minimum width=10mm, minimum height=4mm, fill=red!20, draw=red!80, thick}}
\tikzset{masterlaser/.style={rectangle, minimum width=10mm, minimum height=4mm, fill=green!20, draw=green, thick}}
\tikzset{arrow/.style={line width=1mm}}
\tikzset{node distance = 0mm}
\coordinate (O) at (0,0); 
\coordinate (labelPosition) at (0,55mm); 
\draw[name path = sc1p, spacecraft] (0,0) circle [radius=\spacecraftRadius]
node (sc1) {};
\draw[name path = sc2p, spacecraft] (120:\spacecraftSeparation) circle [radius=\spacecraftRadius]
node (sc2) {};
\draw[name path = sc3p, spacecraft] (60:\spacecraftSeparation) circle [radius=\spacecraftRadius]
node (sc3) {};
\node (sc1l) at($(sc1.center)+(0,5mm)$) {\textbf{SC 1}};
\node (sc2l) at($(sc2.center)+(3mm,-1mm)$) {\textbf{SC 2}};
\node (sc3l) at($(sc3.center)+(-3mm,-1mm)$) {\textbf{SC 3}};
\coordinate (O1) at ($(sc1.center)+(0,-10mm)$); 
\coordinate (O2) at ($(sc2.center)+(150:10mm)$); 
\coordinate (O3) at ($(sc3.center)+(30:10mm)$); 
\path[name path = sc12] (O1) -- (O2);
\path[name path = sc13] (O1) -- (O3);
\path[name path = sc23] (O2) -- (O3);
\draw[name intersections={of=sc12 and sc1p, by=x}] node[laser, rotate=-60] (la12) at(x) {\laserOneTwo};
\draw[name intersections={of=sc12 and sc2p, by=x}] node[laser, rotate=-60] (la21) at(x) {\laserTwoOne};
\draw[name intersections={of=sc13 and sc1p, by=x}] node[laser, rotate=+60] (la13) at(x) {\laserOneThree};
\draw[name intersections={of=sc13 and sc3p, by=x}] node[laser, rotate=+60] (la31) at(x) {\laserThreeOne};
\draw[name intersections={of=sc23 and sc2p, by=x}] node[laser, rotate=  0] (la23) at(x) {\laserTwoThree};
\draw[name intersections={of=sc23 and sc3p, by=x}] node[masterlaser, rotate=  0] (la32) at(x) {\laserThreeTwo};
\draw[-latex, arrow, lred] (la12)--(la21); 
\draw[latex-, arrow, lred] (la13)--(la31); 
\draw[latex-, arrow, lred] (la12.east)..controls +(0.75,-1) and +(-0.75,-1)..(la13.west); 
\draw[-latex, arrow, lred] (la21.west)..controls +(-0.5,1.20) and +(-1.20,0.0)..(la23.west); 
\draw[latex-, arrow, lred] (la31.east)..controls +(0.5,1.20) and +(1.20,0.0)..(la32.east); 
\node[] at (0, 2) {\large \textbf{N5-L32}};
\end{tikzpicture}
\end{minipage}
\hfill
\begin{minipage}{0.3\textwidth}
\begin{alignat*}{3}
&L_{32} &&= L_{32} \\
&L_{23} &&= L_{21} + O_{1} \\
&L_{13} &&= L_{31} + O_{2} + D_{2} \\
&L_{31} &&= L_{32} + O_{3} \\
&L_{21} &&= L_{12} + O_{4} + D_{3} \\
&L_{12} &&= L_{13} + O_{5} \\
\midrule
&L_{32} &&= L_{32} \\
&L_{23} &&= D_{2} + D_{3} + L_{32} + \Sigma_{i=1}^{5} O_{i} \\
&L_{13} &&= D_{2} + L_{32} + O_{2} + O_{3} \\
&L_{31} &&= L_{32} + O_{3} \\
&L_{21} &&= D_{2} + D_{3} + L_{32} + \Sigma_{i=2}^{5} O_{i} \\
&L_{12} &&= D_{2} + L_{32} + O_{2} + O_{3} + O_{5} 
\end{alignat*}
\end{minipage}
\begin{minipage}{0.3\textwidth}
\begin{center}
\vspace{5mm}
\renewcommand{\arraystretch}{1.5}
\begin{tabular}{c | c c c c c c c c }
~ & $D_{1}$ & $D_{2}$ & $D_{3}$ & $O_{1}$ & $O_{2}$ & $O_{3}$ & $O_{4}$ & $O_{5}$ \\
\hline
$B_{22}$ & 0 & 0 & 0 & -1 & 0 & 0 & 0 & 0 \\
$B_{13}$ & 0 & 0 & 0 & 0 & -1 & 0 & 0 & 0 \\
$B_{33}$ & 0 & 0 & 0 & 0 & 0 & -1 & 0 & 0  \\
$B_{21}$ & 0 & 0 & 0 & 0 & 0 & 0 & -1 & 0  \\
$B_{11}$ & 0 & 0 & 0 & 0 & 0 & 0 & 0 & -1 \\
$B_{12}$ & 0 & 0 & 2 & 0 & 0 & 0 & 1 & 0  \\
$B_{23}$ & 1 & -1 & -1 & -1 & -1 & -1 & -1 & -1 \\
$B_{31}$ & 0 & 2 & 0 & 0 & 1 & 0 & 0 & 0  \\
$B_{32}$ & 1 & 1 & 1 & 1 & 1 & 1 & 1 & 1  \\
\end{tabular}
\end{center}
\end{minipage}
\caption{Configuration N5-L32, with three local locks and two inter-spacecraft locks.}
\label{figure:N5-L32}
\end{subfigure}

\begin{subfigure}{\textwidth}
\centering
\begin{minipage}{0.3\textwidth}
\centering
\begin{tikzpicture}
\definecolor{dkblue}{rgb} {0.00,0.33,0.68}
\definecolor{dkred}{rgb} {0.80,0.20,0.00}
\definecolor{lblue}{rgb} {0.00,0.60,0.95}
\definecolor{lred}{rgb} {0.95,0.30,0.00}
\tikzset{spacecraft/.style={circle, minimum width = 24mm, draw=black!60, fill=gray!10, very thick}}
\tikzset{laser/.style={rectangle, minimum width=10mm, minimum height=4mm, fill=red!20, draw=red!80, thick}}
\tikzset{masterlaser/.style={rectangle, minimum width=10mm, minimum height=4mm, fill=green!20, draw=green, thick}}
\tikzset{arrow/.style={line width=1mm}}
\tikzset{node distance = 0mm}
\coordinate (O) at (0,0); 
\coordinate (labelPosition) at (0,55mm); 
\draw[name path = sc1p, spacecraft] (0,0) circle [radius=\spacecraftRadius]
node (sc1) {};
\draw[name path = sc2p, spacecraft] (120:\spacecraftSeparation) circle [radius=\spacecraftRadius]
node (sc2) {};
\draw[name path = sc3p, spacecraft] (60:\spacecraftSeparation) circle [radius=\spacecraftRadius]
node (sc3) {};
\node (sc1l) at($(sc1.center)+(0,5mm)$) {\textbf{SC 1}};
\node (sc2l) at($(sc2.center)+(3mm,-1mm)$) {\textbf{SC 2}};
\node (sc3l) at($(sc3.center)+(-3mm,-1mm)$) {\textbf{SC 3}};
\coordinate (O1) at ($(sc1.center)+(0,-10mm)$); 
\coordinate (O2) at ($(sc2.center)+(150:10mm)$); 
\coordinate (O3) at ($(sc3.center)+(30:10mm)$); 
\path[name path = sc12] (O1) -- (O2);
\path[name path = sc13] (O1) -- (O3);
\path[name path = sc23] (O2) -- (O3);
\draw[name intersections={of=sc12 and sc1p, by=x}] node[laser, rotate=-60] (la12) at(x) {\laserOneTwo};
\draw[name intersections={of=sc12 and sc2p, by=x}] node[laser, rotate=-60] (la21) at(x) {\laserTwoOne};
\draw[name intersections={of=sc13 and sc1p, by=x}] node[laser, rotate=+60] (la13) at(x) {\laserOneThree};
\draw[name intersections={of=sc13 and sc3p, by=x}] node[laser, rotate=+60] (la31) at(x) {\laserThreeOne};
\draw[name intersections={of=sc23 and sc2p, by=x}] node[laser, rotate=  0] (la23) at(x) {\laserTwoThree};
\draw[name intersections={of=sc23 and sc3p, by=x}] node[masterlaser, rotate=  0] (la32) at(x) {\laserThreeTwo};
\draw[latex-, arrow, lblue] (la12)--(la21); 
\draw[latex-, arrow, lblue] (la23)--(la32); 
\draw[-latex, arrow, lblue] (la13)--(la31); 
\draw[-latex, arrow, lblue] (la12.east)..controls +(0.75,-1) and +(-0.75,-1)..(la13.west); 
\draw[latex-, arrow, lblue] (la21.west)..controls +(-0.5,1.20) and +(-1.20,0.0)..(la23.west); 
\node[] at (0, 2) {\large \textbf{N6-L32}};
\end{tikzpicture}
\end{minipage}
\hfill
\begin{minipage}{0.3\textwidth}
\begin{alignat*}{3}
&L_{32} &&= L_{32} \\
&L_{23} &&= L_{32} + O_{1} + D_{1} \\
&L_{13} &&= L_{12} + O_{2} \\
&L_{31} &&= L_{13} + O_{3} + D_{2} \\
&L_{21} &&= L_{23} + O_{4} \\
&L_{12} &&= L_{21} + O_{5} + D_{3} \\
\midrule
&L_{32} &&= L_{32} \\
&L_{23} &&= D_{1} + L_{32} + O_{1} \\
&L_{13} &&= D_{1} + D_{3} + L_{32} + O_{1} + O_{2} + O_{4} + O_{5} \\
&L_{31} &&= D_{1} + D_{2} + D_{3} + L_{32} + \Sigma_{i=1}^{5} O_{i} \\
&L_{21} &&= D_{1} + L_{32} + O_{1} + O_{4} \\
&L_{12} &&= D_{1} + D_{3} + L_{32} + O_{1} + O_{4} + O_{5} 
\end{alignat*}
\end{minipage}
\begin{minipage}{0.3\textwidth}
\begin{center}
\vspace{5mm}
\renewcommand{\arraystretch}{1.5}
\begin{tabular}{c | c c c c c c c c}
~ & $D_{1}$ & $D_{2}$ & $D_{3}$ & $O_{1}$ & $O_{2}$ & $O_{3}$ & $O_{4}$ & $O_{5}$ \\
\hline
$B_{23}$ & 0 & 0 & 0 & -1 & 0 & 0 & 0 & 0 \\
$B_{11}$ & 0 & 0 & 0 & 0 & 1 & 0 & 0 & 0 \\
$B_{31}$ & 0 & 0 & 0 & 0 & 0 & -1 & 0 & 0 \\
$B_{22}$ & 0 & 0 & 0 & 0 & 0 & 0 & 1 & 0 \\
$B_{12}$ & 0 & 0 & 0 & 0 & 0 & 0 & 0 & -1 \\
$B_{13}$ & 0 & 2 & 0 & 0 & 0 & 1 & 0 & 0 \\
$B_{21}$ & 0 & 0 & 2 & 0 & 0 & 0 & 0 & 1 \\
$B_{32}$ & 2 & 0 & 0 & 1 & 0 & 0 & 0 & 0 \\
$B_{33}$ & -1 & -1 & -1 & -1 & -1 & -1 & -1 & -1\\
\end{tabular}
\end{center}
\end{minipage}
\caption{Configuration N6-L32, with two local locks and three inter-spacecraft locks.}
\label{figure:N6-L32}
\end{subfigure}
\caption{Laser locking configurations N4-L32, N5-L32, and N6-L32, with LA 32 as the primary laser.}
\label{figure:N4-N6-L32}
\end{figure*}


\section{Results} \label{section:results}

The result of the preceding procedure consists of a catalog of valid locking schemes for every possible fixed primary laser. All six valid configurations for the primary laser L32 in the non-swap scheme are shown in Figures~\ref{figure:N1-N3-L32} and \ref{figure:N4-N6-L32}. The complete catalogue accompanies this work as a supplementary document. 

In these figures, for each configuration, the top set of equations are the initial locking scheme dependencies that reflect how each laser is phase-locked to another feasible one; the bottom set of equations are the symbolic solution of this system of equations which yields the secondary laser frequencies in terms of the primary laser plus the combined effect of the transponder offsets and Doppler shifts accumulated due to long-arm propagation. 

The matrices to the right represent the coefficients of the linear dependencies of the resulting beatnotes with respect to the Doppler shifts and offset frequencies. Notice that there are always five beatnotes that correspond, up to a sign, to the offsets themselves, representing the locking beatnotes, whereas the other four beatnotes represent non-locking beatnotes. This distinction plays a role in the frequency plan due to the nature of the constraints on the beatnotes. 

The non-swap configurations are labeled N1 to N6, followed by the designation of primary laser (e.g., `L12' if the primary laser is located in spacecraft 1 and pointed to spacecraft 2). The frequency swap configurations are, in turn, labeled S1 to S12, followed by the designation of primary laser. The sorting of the configurations from 1 to 6 in the non-swap case, or from 1 to 12 in the frequency-swap case, for a given choice of primary laser, is done by ascending level of complexity.

The complexity of each locking scheme is quantified using two key metrics: 
\begin{enumerate}
	\item the sum of ``distances'' from the primary laser to all other lasers in the locking network, and
	\item the total number of local locks.
\end{enumerate}
Here, distance refers not to physical separation in space, but to the relative position of a laser within the phase-locking network, measured as the number of phase-locking steps being taken to propagate the frequency stability of the primary laser to it.

To compute these distances, we construct a graph representation of each locking scheme, where lasers serve as nodes and phase-locking dependencies define the edges (see, e.g., Figure~\ref{figure:N1L12-S1L12}). We then apply a Breadth-First Search (BFS) algorithm to determine the shortest path from the primary laser to all other lasers in the scheme. Finally, the distances are summed, resulting in a number that reflects the amount of phase-locking steps required to propagate coherence throughout the network. For those configurations that result in the same sum of distances, we use the count of local locks (i.e., instances where there is a direct phase-lock between two local lasers), as a secondary criterion for sorting.  

Local locks contribute to the stability of the phase-locking architecture by reducing reliance on inter-spacecraft links, which introduce Doppler-shifts and rely on interferometric readout under weak light conditions. Configurations with a higher number of local locks are favored when the sum of distances is identical. By establishing more direct links between co-located lasers, these configurations ensure a more robust transfer of frequency stability throughout the network.

By systematically ranking all valid configurations for each choice of primary laser, we establish a structured hierarchy of laser-locking schemes. The resulting classification not only provides insight into the internal dependencies of each scheme, but can also inform the selection of robust frequency plans under any possible operational scenario.

\section{Conclusion} \label{section:conclusion}

We have developed a method to systematically discover, validate, and classify all laser-locking schemes for LISA. The outcome is a complete catalog of valid laser transponder configurations, including the linear equations relating the laser frequencies to a primary laser, and the corresponding dependencies between the beatnotes and the offsets and Doppler shifts. This work thus complements the frequency planning algorithm for LISA by providing the valid set of possible schemes; choosing one plays a fundamental role in determining the constraints of the optimization method that generates the frequency offsets needed to compensate for the Doppler shifts during orbit, such that all beatnote frequencies stay in-band throughout the mission lifetime.

\section{References}

\begin{thebibliography}{9}%
\makeatletter
\providecommand \@ifxundefined [1]{%
 \@ifx{#1\undefined}
}%
\providecommand \@ifnum [1]{%
 \ifnum #1\expandafter \@firstoftwo
 \else \expandafter \@secondoftwo
 \fi
}%
\providecommand \@ifx [1]{%
 \ifx #1\expandafter \@firstoftwo
 \else \expandafter \@secondoftwo
 \fi
}%
\providecommand \natexlab [1]{#1}%
\providecommand \enquote  [1]{``#1''}%
\providecommand \bibnamefont  [1]{#1}%
\providecommand \bibfnamefont [1]{#1}%
\providecommand \citenamefont [1]{#1}%
\providecommand \href@noop [0]{\@secondoftwo}%
\providecommand \href [0]{\begingroup \@sanitize@url \@href}%
\providecommand \@href[1]{\@@startlink{#1}\@@href}%
\providecommand \@@href[1]{\endgroup#1\@@endlink}%
\providecommand \@sanitize@url [0]{\catcode `\\12\catcode `\$12\catcode `\&12\catcode `\#12\catcode `\^12\catcode `\_12\catcode `\%12\relax}%
\providecommand \@@startlink[1]{}%
\providecommand \@@endlink[0]{}%
\providecommand \url  [0]{\begingroup\@sanitize@url \@url }%
\providecommand \@url [1]{\endgroup\@href {#1}{\urlprefix }}%
\providecommand \urlprefix  [0]{URL }%
\providecommand \Eprint [0]{\href }%
\providecommand \doibase [0]{http://dx.doi.org/}%
\providecommand \selectlanguage [0]{\@gobble}%
\providecommand \bibinfo  [0]{\@secondoftwo}%
\providecommand \bibfield  [0]{\@secondoftwo}%
\providecommand \translation [1]{[#1]}%
\providecommand \BibitemOpen [0]{}%
\providecommand \bibitemStop [0]{}%
\providecommand \bibitemNoStop [0]{.\EOS\space}%
\providecommand \EOS [0]{\spacefactor3000\relax}%
\providecommand \BibitemShut  [1]{\csname bibitem#1\endcsname}%
\let\auto@bib@innerbib\@empty
\bibitem [{\citenamefont {Miller}\ and\ \citenamefont {Yunes}(2019)}]{Miller2019}%
  \BibitemOpen
  \bibfield  {author} {\bibinfo {author} {\bibfnamefont {M.~C.}\ \bibnamefont {Miller}}\ and\ \bibinfo {author} {\bibfnamefont {N.}~\bibnamefont {Yunes}},\ }\href {\doibase 10.1038/s41586-019-1129-z} {\bibfield  {journal} {\bibinfo  {journal} {Nature}\ }\textbf {\bibinfo {volume} {568}},\ \bibinfo {pages} {469} (\bibinfo {year} {2019})}\BibitemShut {NoStop}%
\bibitem [{\citenamefont {Collaboration}(2015)}]{Collaboration2015}%
  \BibitemOpen
  \bibfield  {author} {\bibinfo {author} {\bibfnamefont {L.~S.}\ \bibnamefont {Collaboration}},\ }\href {\doibase 10.1088/0264-9381/32/7/074001} {\bibfield  {journal} {\bibinfo  {journal} {Class. Quantum Grav.}\ }\textbf {\bibinfo {volume} {32}},\ \bibinfo {pages} {074001} (\bibinfo {year} {2015})}\BibitemShut {NoStop}%
\bibitem [{\citenamefont {Acernese}\ \emph {et~al.}(2014)\citenamefont {Acernese}, \citenamefont {Agathos}, \citenamefont {Agatsuma}, \citenamefont {Aisa}, \citenamefont {Allemandou}, \citenamefont {Allocca}, \citenamefont {Amarni}, \citenamefont {Astone}, \citenamefont {Balestri}, \citenamefont {Ballardin} \emph {et~al.}}]{AdvancedVirgo15}%
  \BibitemOpen
  \bibfield  {author} {\bibinfo {author} {\bibfnamefont {F.}~\bibnamefont {Acernese}}, \bibinfo {author} {\bibfnamefont {M.}~\bibnamefont {Agathos}}, \bibinfo {author} {\bibfnamefont {K.}~\bibnamefont {Agatsuma}}, \bibinfo {author} {\bibfnamefont {D.}~\bibnamefont {Aisa}}, \bibinfo {author} {\bibfnamefont {N.}~\bibnamefont {Allemandou}}, \bibinfo {author} {\bibfnamefont {A.}~\bibnamefont {Allocca}}, \bibinfo {author} {\bibfnamefont {J.}~\bibnamefont {Amarni}}, \bibinfo {author} {\bibfnamefont {P.}~\bibnamefont {Astone}}, \bibinfo {author} {\bibfnamefont {G.}~\bibnamefont {Balestri}}, \bibinfo {author} {\bibfnamefont {G.}~\bibnamefont {Ballardin}},  \emph {et~al.},\ }\href {\doibase 10.1088/0264-9381/32/2/024001} {\bibfield  {journal} {\bibinfo  {journal} {Classical and Quantum Gravity}\ }\textbf {\bibinfo {volume} {32}},\ \bibinfo {pages} {024001} (\bibinfo {year} {2014})}\BibitemShut {NoStop}%
\bibitem [{\citenamefont {Abbott}\ \emph {et~al.}(2016)\citenamefont {Abbott}, \citenamefont {Abbott}, \citenamefont {Abbott}, \citenamefont {Abernathy}, \citenamefont {Acernese}, \citenamefont {Ackley}, \citenamefont {Adams}, \citenamefont {Adams}, \citenamefont {Addesso}, \citenamefont {Adhikari} \emph {et~al.}}]{Abbott2016}%
  \BibitemOpen
  \bibfield  {author} {\bibinfo {author} {\bibfnamefont {B.~P.}\ \bibnamefont {Abbott}}, \bibinfo {author} {\bibfnamefont {R.}~\bibnamefont {Abbott}}, \bibinfo {author} {\bibfnamefont {T.~D.}\ \bibnamefont {Abbott}}, \bibinfo {author} {\bibfnamefont {M.~R.}\ \bibnamefont {Abernathy}}, \bibinfo {author} {\bibfnamefont {F.}~\bibnamefont {Acernese}}, \bibinfo {author} {\bibfnamefont {K.}~\bibnamefont {Ackley}}, \bibinfo {author} {\bibfnamefont {C.}~\bibnamefont {Adams}}, \bibinfo {author} {\bibfnamefont {T.}~\bibnamefont {Adams}}, \bibinfo {author} {\bibfnamefont {P.}~\bibnamefont {Addesso}}, \bibinfo {author} {\bibfnamefont {R.~X.}\ \bibnamefont {Adhikari}},  \emph {et~al.} (\bibinfo {collaboration} {LIGO Scientific Collaboration and Virgo Collaboration}),\ }\href {\doibase 10.1103/PhysRevLett.116.061102} {\bibfield  {journal} {\bibinfo  {journal} {Phys. Rev. Lett.}\ }\textbf {\bibinfo {volume} {116}},\ \bibinfo {pages} {061102} (\bibinfo {year} {2016})}\BibitemShut {NoStop}%
\bibitem [{\citenamefont {Abbott}\ \emph {et~al.}(2017)\citenamefont {Abbott}, \citenamefont {Abbott}, \citenamefont {Abbott}, \citenamefont {Acernese}, \citenamefont {Ackley}, \citenamefont {Adams}, \citenamefont {Adams}, \citenamefont {Addesso}, \citenamefont {Adhikari}, \citenamefont {Adya} \emph {et~al.}}]{Abbott2017}%
  \BibitemOpen
  \bibfield  {author} {\bibinfo {author} {\bibfnamefont {B.~P.}\ \bibnamefont {Abbott}}, \bibinfo {author} {\bibfnamefont {R.}~\bibnamefont {Abbott}}, \bibinfo {author} {\bibfnamefont {T.~D.}\ \bibnamefont {Abbott}}, \bibinfo {author} {\bibfnamefont {F.}~\bibnamefont {Acernese}}, \bibinfo {author} {\bibfnamefont {K.}~\bibnamefont {Ackley}}, \bibinfo {author} {\bibfnamefont {C.}~\bibnamefont {Adams}}, \bibinfo {author} {\bibfnamefont {T.}~\bibnamefont {Adams}}, \bibinfo {author} {\bibfnamefont {P.}~\bibnamefont {Addesso}}, \bibinfo {author} {\bibfnamefont {R.~X.}\ \bibnamefont {Adhikari}}, \bibinfo {author} {\bibfnamefont {V.~B.}\ \bibnamefont {Adya}},  \emph {et~al.} (\bibinfo {collaboration} {LIGO Scientific Collaboration and Virgo Collaboration}),\ }\href {\doibase 10.1103/PhysRevLett.119.161101} {\bibfield  {journal} {\bibinfo  {journal} {Phys. Rev. Lett.}\ }\textbf {\bibinfo {volume} {119}},\ \bibinfo {pages} {161101} (\bibinfo {year} {2017})}\BibitemShut {NoStop}%
\bibitem [{\citenamefont {Abbott}\ \emph {et~al.}(2021)\citenamefont {Abbott}, \citenamefont {Abbott}, \citenamefont {Abraham}, \citenamefont {Acernese}, \citenamefont {Ackley}, \citenamefont {Adams}, \citenamefont {Adams}, \citenamefont {Adhikari}, \citenamefont {Adya}, \citenamefont {Affeldt} \emph {et~al.}}]{Abbott2021}%
  \BibitemOpen
  \bibfield  {author} {\bibinfo {author} {\bibfnamefont {R.}~\bibnamefont {Abbott}}, \bibinfo {author} {\bibfnamefont {T.~D.}\ \bibnamefont {Abbott}}, \bibinfo {author} {\bibfnamefont {S.}~\bibnamefont {Abraham}}, \bibinfo {author} {\bibfnamefont {F.}~\bibnamefont {Acernese}}, \bibinfo {author} {\bibfnamefont {K.}~\bibnamefont {Ackley}}, \bibinfo {author} {\bibfnamefont {A.}~\bibnamefont {Adams}}, \bibinfo {author} {\bibfnamefont {C.}~\bibnamefont {Adams}}, \bibinfo {author} {\bibfnamefont {R.~X.}\ \bibnamefont {Adhikari}}, \bibinfo {author} {\bibfnamefont {V.~B.}\ \bibnamefont {Adya}}, \bibinfo {author} {\bibfnamefont {C.}~\bibnamefont {Affeldt}},  \emph {et~al.},\ }\href {\doibase 10.3847/2041-8213/ac082e} {\bibfield  {journal} {\bibinfo  {journal} {The Astrophysical Journal Letters}\ }\textbf {\bibinfo {volume} {915}},\ \bibinfo {pages} {L5} (\bibinfo {year} {2021})}\BibitemShut {NoStop}%
\bibitem [{\citenamefont {Cahillane}\ and\ \citenamefont {Mansell}(2022)}]{Cahillane2022}%
  \BibitemOpen
  \bibfield  {author} {\bibinfo {author} {\bibfnamefont {C.}~\bibnamefont {Cahillane}}\ and\ \bibinfo {author} {\bibfnamefont {G.}~\bibnamefont {Mansell}},\ }\href {\doibase 10.3390/galaxies10010036} {\bibfield  {journal} {\bibinfo  {journal} {Galaxies}\ }\textbf {\bibinfo {volume} {10}},\ \bibinfo {pages} {36} (\bibinfo {year} {2022})}\BibitemShut {NoStop}%
\bibitem [{\citenamefont {Amaro-Seoane}\ \emph {et~al.}(2023)\citenamefont {Amaro-Seoane}, \citenamefont {Andrews}, \citenamefont {Arca~Sedda}, \citenamefont {Askar}, \citenamefont {Baghi}, \citenamefont {Balasov}, \citenamefont {Bartos}, \citenamefont {Bavera}, \citenamefont {Bellovary}, \citenamefont {Berry} \emph {et~al.}}]{Amaro2023}%
  \BibitemOpen
  \bibfield  {author} {\bibinfo {author} {\bibfnamefont {P.}~\bibnamefont {Amaro-Seoane}}, \bibinfo {author} {\bibfnamefont {J.}~\bibnamefont {Andrews}}, \bibinfo {author} {\bibfnamefont {M.}~\bibnamefont {Arca~Sedda}}, \bibinfo {author} {\bibfnamefont {A.}~\bibnamefont {Askar}}, \bibinfo {author} {\bibfnamefont {Q.}~\bibnamefont {Baghi}}, \bibinfo {author} {\bibfnamefont {R.}~\bibnamefont {Balasov}}, \bibinfo {author} {\bibfnamefont {I.}~\bibnamefont {Bartos}}, \bibinfo {author} {\bibfnamefont {S.~S.}\ \bibnamefont {Bavera}}, \bibinfo {author} {\bibfnamefont {J.}~\bibnamefont {Bellovary}}, \bibinfo {author} {\bibfnamefont {C.~P.~L.}\ \bibnamefont {Berry}},  \emph {et~al.},\ }\href {\doibase 10.1007/s41114-022-00041-y} {\bibfield  {journal} {\bibinfo  {journal} {Living Reviews in Relativity}\ }\textbf {\bibinfo {volume} {26}} (\bibinfo {year} {2023}),\ 10.1007/s41114-022-00041-y}\BibitemShut {NoStop}%
\bibitem [{\citenamefont {Heinzel}\ \emph {et~al.}(2024)\citenamefont {Heinzel}, \citenamefont {\'Alvarez-Vizoso}, \citenamefont {Dovale-\'Alvarez},\ and\ \citenamefont {Wiesner}}]{Heinzel2024}%
  \BibitemOpen
  \bibfield  {author} {\bibinfo {author} {\bibfnamefont {G.}~\bibnamefont {Heinzel}}, \bibinfo {author} {\bibfnamefont {J.}~\bibnamefont {\'Alvarez-Vizoso}}, \bibinfo {author} {\bibfnamefont {M.}~\bibnamefont {Dovale-\'Alvarez}}, \ and\ \bibinfo {author} {\bibfnamefont {K.}~\bibnamefont {Wiesner}},\ }\href {\doibase 10.1103/PhysRevD.110.042002} {\bibfield  {journal} {\bibinfo  {journal} {Phys. Rev. D}\ }\textbf {\bibinfo {volume} {110}},\ \bibinfo {pages} {042002} (\bibinfo {year} {2024})}\BibitemShut {NoStop}%
\end{thebibliography}
%


\section*{Acknowledgements}

The authors acknowledge financial support by the German Aerospace Center (DLR) with funds from the Federal Ministry of Economics and Technology (BMWi) according to a decision of the German Federal Parliament (Grants No.\ 50OQ0601, No.\ 50OQ1301, No.\ 50OQ1801), the European Space Agency (ESA) (Grants No.\ 22331/09/ NL/HB, No.\ 16238/10/NL/HB), the Deutsche Forschungsgemeinschaft (DFG) Sonderforschungsbereich 1128 Relativistic Geodesy and Cluster of Excellence ``QuantumFrontiers: Light and Matter at the Quantum Frontier: Foundations and Applications in Metrology'' (EXC-2123, Project No.\ 390837967). Furthermore, this work was supported by the LEGACY cooperation on low-frequency gravitational wave astronomy (M.IF.A.QOP18098).


\end{document}